\documentclass[aps,column,amsfonts,amssymb,superscriptaddress,longbibliography]{revtex4}
\usepackage{graphicx}
\usepackage{epstopdf} 
\usepackage{dcolumn}
\usepackage{bm}
\usepackage{bbm,bbold}
\usepackage{color}
\usepackage{xcolor, soul}
\sethlcolor{green}
\usepackage{amsfonts}
\usepackage{amsmath}
\usepackage{mdframed}
\usepackage[normalem]{ulem}
\usepackage{mathrsfs}   
\usepackage[none]{hyphenat}
\usepackage{subfigure}
\usepackage{float}
\usepackage{physics}
\usepackage[colorlinks=true,citecolor=blue]{hyperref}
\hypersetup{colorlinks=true,citecolor=blue,linkcolor=blue,urlcolor=blue}

\begin{document}
\title{Light-induced Nonlinear Spin Hall Current in Single-layer WTe$_2$} 
\author{Pankaj Bhalla}
\affiliation{Department of Physics, School of Engineering and Sciences, SRM University AP, Amaravati, 522240, India}
\author{Habib Rostami}
\affiliation{Department of Physics, University of Bath, Claverton Down, Bath BA2 7AY, United Kingdom}
\affiliation{Nordita, KTH Royal Institute of Technology and Stockholm University,
Hannes Alfv\'ens v\"ag 12, 10691 Stockholm, Sweden}
\date{\today}
\begin{abstract}
In this theoretical investigation, we analyze light-induced nonlinear spin Hall currents in a gated single-layer 1T$'$-WTe$_2$, flowing transversely to the incident laser polarization direction. Our study encompasses the exploration of the second and third-order rectified spin Hall currents using an effective low-energy Hamiltonian and employing the Kubo's formalism.
We extend our analysis to a wide frequency range spanning both transparent and absorbing regimes, investigating the influence of light frequency below and above the optical band gap.
Additionally, we investigate the influence of an out-of-plane gate potential on the system, disrupting inversion symmetry and effectively manipulating both the strength and sign of nonlinear spin Hall responses.  We predict a pronounced third-order spin Hall current relative to its second-order counterpart. The predicted nonlinear spin currents show strong anisotropic dependence on the laser polarization angle.
The outcomes of our study contribute to a generalized framework for nonlinear response theory within the spin channel will impact the development of emerging field of opto-spintronic.
\end{abstract}
\maketitle
\section{Introduction}

Electronic transport in two-dimensional (2D) quantum materials has emerged as a central topic for new physics and novel technologies in the spintronic industry and future energy-saving devices~\cite{keimer_NP2017, Giustino_2021JPM}. The second-order nonlinear photocurrent in novel quantum materials is attracting interest from both applied and fundamental point of view~\cite{Sodemann_prl_2015,Takahiro_sciadv_2016,deJuan2017,habib_PRB2018,Hua_sciadv_2019,mueller_NPJ2D2018, soavi_nn_2018,Principi_prb_2019, bhalla_PRL2020, Wang_AdM_2021, klimmer_NP2021, bhalla_PRB2021, sinha_NP2022, bhalla_PRB2022, chakraborty_2Dmat2022, bhalla_PRL2022, lahiri_arxiv2022, bhalla_PRL2021,zeng_PRB2022, bhalla_PRB2023a, bhalla_PRB2022Er, ikeda_arxiv2023}. In particular, one of the main aims is to realize topological protection of nonlinear current in topological quantum materials such as Dirac and Weyl systems~\cite{deJuan2017,habib_PRB2018,shvetsov_JETP2019, tiwari_2021}.
Experimental measurements in 2D WTe$_2$ show that the nonlinear Hall effect arises due to the momentum derivative of the Berry curvature, so-called the Berry curvature dipole~\cite{Sodemann_prl_2015} that can be tuned with the out-of-plane potential bias via the top and bottom gate voltages~\cite{ma_Nat2019, kang_NM2019}. The counter propagation of electrons with opposite spins~\cite{sun_PRB2005} conceives the celebrated spin Hall effect on applying a bias voltage between two contacts~\cite{kane_PRL2005,bernevig_Sci2006, bernevig_PRL2006, qi_PRB2006}. The phenomenon arises due to the strong spin-orbit coupling, a key element in their electronic band structure, and generating a spin current~\cite{bernevig_Sci2006, xu_PRB2006}. 
A generalization to the nonlinear regime is a rapidly growing field of research. For instance, we recall theoretical and experimental studies on photoinduced second-order spin current in quantum materials with spin-orbit coupling \cite{Ishizuka_prl_2022,Sakimura2014,Nair_prb_2020,Hamamoto_prb_2017,Bhat_prl_2005,Lihm_prb_2022,Hua_sciadv_2019}.  

 \begin{figure}
    \centering
    \includegraphics[width=10cm]{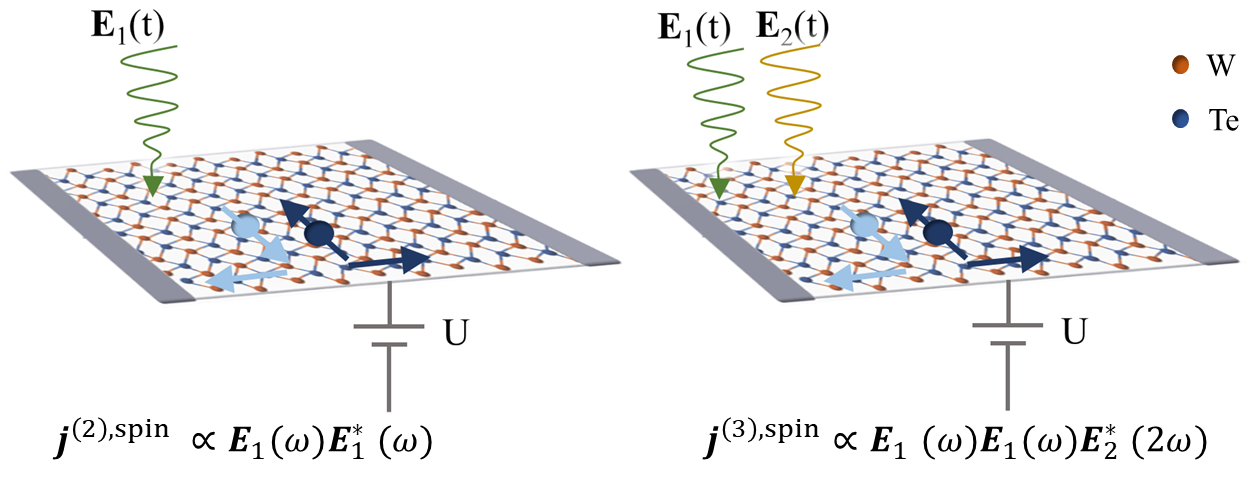}
    \caption{Schematic setups for the light-induced nonlinear rectified spin current in a 2D WTe$_2$ device. The left panel refers to the second-order spin current that arises in response to the single-color light beam. The right panel stems from the interference of two-color light beams and represents the third-order rectified spin current. Here ${\bm E}_1(t)$ and ${\bm E}_2(t)$ refer to the electric fields, and $U$ corresponds to the out-of-plane gate potential.}
    \label{fig:Fig1}
\end{figure}

Measurement of nonlinear Hall effect~\cite{Du_nrp_2021} using infrared pulse excitation, and high-temperature intrinsic spin-Hall effect~\cite{Garcia_prl_2020} in two-dimensional WTe$_2$ motivates to look for the nonlinear generalization of the spin Hall current in this exotic quantum material with a rich ground state phase diagram. The generation of the rectified spin current has been widely studied in traditional semiconductors and quantum well structures, where a linear wave vector term in the Bloch Hamiltonian stemming from the spin-orbit interaction leads to the spin splitting in the band structure and plays the significant role~\cite{zhou_PRB2007, ivchenko_SST2008, tarasenko_JETP2005, zhao_PRB2005, wang_PRL2010, hamamoto_PRB2017}. The nonlinear spin current has yet to be discussed or explored for monolayer WTe$_2$ having different crystalline symmetry, which is the main aim of the present work. The time-reversal 1T$'$-WTe$_2$ structure yields second-order current when the inversion symmetry is broken~\cite{bhalla_PRB2022,du_PRL2018} by an out-of-plane potential bias that allows tuning the gap associated with each valley and spin~\cite{xu_NP2018}. This results in a phase transition driven by the interplay of the spin-orbit coupling and the out-of-plane bias~\cite{qian_Sc2014}.

The second-order rectified current $j^{(2)}_{dc} \sim E(\omega)E^\ast(\omega)$ can be generated in non-centrosymmetric materials. There is, however, a growing interest in the third-order rectified current that can be achieved in the presence of an extra direct electric field $j^{(3)}_{dc} \sim E_{dc} E(\omega)E^\ast(\omega)$ which can induce rectified current in centrosymmetric systems \cite{Fregoso_prb_2019, fregoso_PRL2018, Ahn2022}. 
Another third-order rectification process is obtained in response to a two-color driving field.
The two-color field is formally written as 
\begin{align}
 {\bm E}(t) &= {\bm E}_1 e^{i(\omega_1 t+\phi_1)} + {\bm E}_2  e^{i(\omega_2 t+\phi_2)} +c.c.~,
\end{align}
where ${\bm E}_1$, and ${\bm E}_2$ are the electric fields associated with frequency $\omega_1$ and $\omega_2$, and $\phi_1$ and $\phi_2$ phases respectively.
Two-color optical rectification is a process that involves the interference of single frequency beam with the second-harmonic laser beam $j^{(3)}_{dc} \sim E_1(\omega_1)  E_1(\omega_1) E^\ast_2(\omega_2)$ with $\omega_2=2\omega_1$ which was first introduced by Manykin and Alfanasev~\cite{manykin_JETP1967}. The process has been extensively exploited in semiconductors and two-dimensional materials by illuminating a combination of monochromatic beams of frequency $\omega$ and $2\omega$, leads to one-photon and two-photon absorption transitions~\cite{bhat_PRL2000, rioux_PRB2011, khurgin_JNOPM1995, totero_PRL2020, cheng_NJP2014, cheng_PRB2015,atanasov_PRL1996, hache_PRL1997}. 

In addition to the interest in the Fermi-surface contribution to the low-frequency (e.g. THz range) photocurrent generation~\cite{Principi_prb_2019,sodemann_PRL2015,Du2021,gao_PRR2021,Zhang_Fu_PNAS_2021}, recent studies have also predicted the presence of photocurrent in the transparent frequency range based on various mechanisms~\cite{kalpan_PRL2020, onishi_PRB2022, golub_PRB2022, shi_PRB2023,Matsyshyn_prb_2023}. This process is highly counter-intuitive since the typical photocurrent generation mechanisms in semiconductors, namely second-order shift and injection photocurrents, require a real optical transition of electrons from the valence to the conduction band. Therefore, the linear optical absorption is accompanied with the conventional photocurrent generation in semiconductors. The rationale behind the generation of photocurrent relies on various mechanisms, including time-reversal symmetry breaking~\cite{kalpan_PRL2020}, Raman scattering of light~\cite{golub_PRB2022}, residual intraband absorption~\cite{shi_PRB2023} with a small but finite carrier relaxation rate~\cite{onishi_PRB2022}. Using the Floquet model, it is shown that the occupation number becomes a ``staircase'' form, allowing for a finite rectified electric current within the optical gap \cite{Matsyshyn_prb_2023}.
It is worth highlighting that the second-order current cannot generate Joule heating, as the time-averaged dissipated power, represented as $W^{(2)}=-\langle {\bm j}^{(2)}(t)\cdot {\bm E}(t) \rangle_{\text{time-average}},$ is always zero for a periodic electric field  \cite{Bassani1991}. This is because, for both the second-harmonic and rectification currents, the transient power density will oscillate in time leading to a vanishing time-averaged value. 

In this article, we theoretically investigate the nonlinear rectified spin current in monolayer WTe$_2$--see Fig.~\ref{fig:Fig1}, based on second-order and third-order response theories. We utilize a many-body diagrammatic perturbation technique. We discuss that the transverse nonlinear charge current vanishes due to the symmetry conditions, but a nonlinear spin current remains finite in both second and third-order rectification mechanisms.  
The current injection mechanism dominantly drives the second-order spin current, which is finite in the interband frequency regime $\hbar\omega>\Delta_{op}$ where $\Delta_{op}$ is the optical gap based on the Pauli exclusion principle at zero electronic temperature. However, the two-color spin current mechanism is more complex with non-trivial features. Intriguingly, we obtain two-color spin response that can be finite in both $\hbar\omega<\Delta_{op}$ and $\hbar\omega>\Delta_{op}$ regimes. In simpler terms, the effect below the gap does not stem from the transport of the photoexcited carrier population; instead, it originates from virtual excitations in transparent regime, akin to the references mentioned \cite{kalpan_PRL2020, onishi_PRB2022, golub_PRB2022, shi_PRB2023, Matsyshyn_prb_2023}.
In the remainder of this paper, we present a quantitative analysis of second and third-order spin susceptibilities and discuss their dependence on laser frequency, external displacement field (gate potential), chemical potential, and laser polarization angle. 

\section{Theory and Method} 
 \begin{figure*}
    \centering
    \includegraphics[width=16cm]{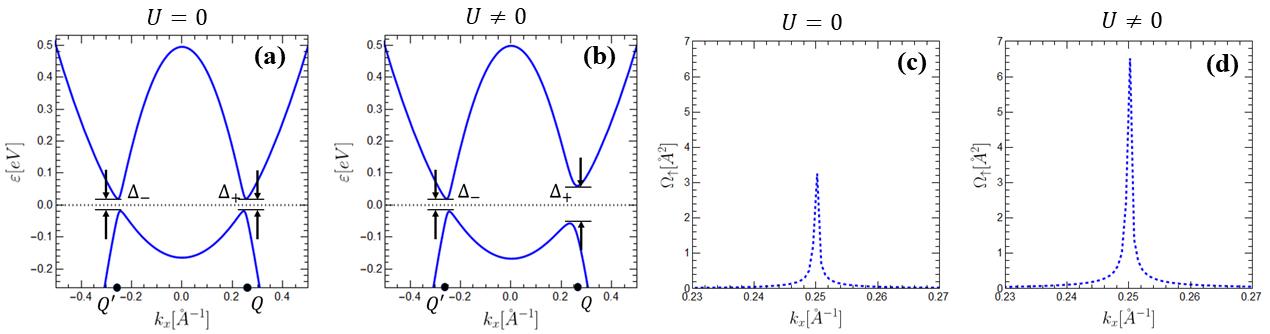}
    \caption{Panel (a) and (b) refers to the spin-up energy dispersion of 1T$'$-WTe$_2$ with the back gate potential $U = 0$ and $U = 2\delta_{{\rm SOC}}$. Here, $\Delta_{\pm} = 2| U \pm \delta_{\rm SOC}|$ is the energy gap around two valley points $Q$ and $Q'$ with $2\delta_{\rm SOC}$ being the spin-orbit gap. The horizontal dotted black line refers to the chemical potential $\mu =0$. Panel (c) and (d) illustrate the spin-up Berry curvature variation for $U=0$ and $U =\delta_{{\rm SOC}}$, respectively.}
    \label{fig:Fig2}
\end{figure*}
We utilize an effective continuum ${\bm k}\cdot {\bm p}$ model Hamiltonian for the 1T$'$ phase of the monolayer WTe$_2$ which incorporates p-orbital of Te and d-orbital of W. This Hamiltonian can nicely describe the low-energy bands, which has the following representation~\cite{xu_NP2018, garcia_PRL2020}:
\begin{align} \label{eqn:hamil}
    \mathcal{\hat H} =& \{A k^2  {\hat \sigma}_0 + (\delta + Bk^2) \hat \sigma_z + v_y k_y \hat \sigma_y 
    + U\hat \sigma_x \} \otimes \hat s_0  + v_x  k_x  \hat \sigma_x \otimes  \hat s_y ,
\end{align}
where $\hat{\sigma}_{i=x,y,z}$ and $\hat{s}_{i=x,y,z}$ are Pauli matrices in the orbital and spin basis, respectively. The $\hat{\sigma}_0$ ($\hat s_0$) is the identity matrix in the orbital (spin) basis. The parameter $2\delta$ stands for the gap at $\Gamma$-point, and $v_y$ characterizes the anisotropy in the momentum space. The parameters $2A =  1/m_p - 1/m_d $ and $2B = 1/m_p + 1/m_d$, where $m_p$ and $m_d$ define the effective masses of p-orbital of Te and d-orbital of W respectively. The parameter $U$ represents the displacement field, the coupling between the out-of-plane electric field and the orbitals which breaks the inversion symmetry of the system, $v_x$ is the spin-orbit coupling strength, and the wave vector ${\bm k} = (k_x,k_y)$ having $k = |{\bm k}|$. The parameter values are considered from the {\em ab initio} band structure calculations~\cite{qian_Sc2014} and fitted with the experimental predictions to obtain the spin-orbit coupling gap $\delta_{{\rm {SOC}}}= v_xQ = 45$ meV at $ Q = \sqrt{|\delta|/2B}$ Dirac points in the momentum space ~\cite{tang_NP2017}.

For the above model Hamiltonian considered, the Berry curvature ($U=0$ case) reads  
\begin{align} \label{eqn:jj}
     \Omega^{c}_{{\bm k},s}= - \Omega^{v}_{{\bm k},s} &= - \frac{s v_x v_y}{N}
    \bigg(h_z (v_x^2 k_x^2 - v_y^2 k_y^2) 
    - |h_{12} |^2 2B(k_x^2 - k_y^2)\bigg),
\end{align}
where $N = |h_{12} | \sqrt{[h_{z}]^2  + [h_{12}]^2}$, $h_{12} = s v_x k_x -iv_y k_y$ and $h_z = \delta + Bk^2$. As seen the Berry curvature is proportional to the spin index, thus flips sign for up and down spin indices. 
The energy dispersion and the Berry curvature corresponding to the spin-up component of the Hamiltonian given in Eq.~\eqref{eqn:hamil} are depicted in Fig.~\ref{fig:Fig2}.

\begin{figure*}
    \centering
    \includegraphics[width=14cm]{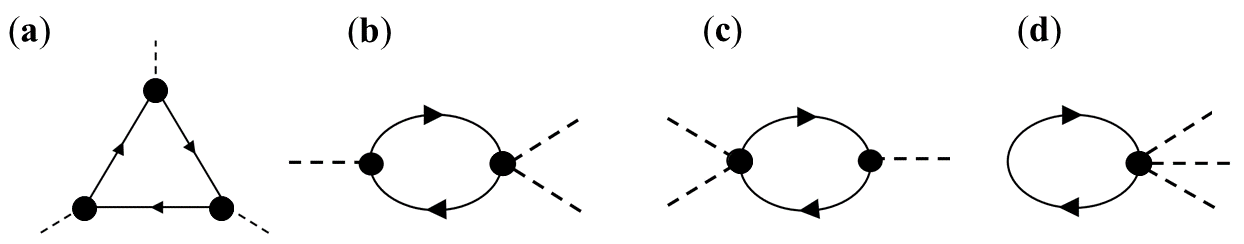}
    \caption{Feynman diagrams for the second-order response. (a) correspond to the paramagnetic contributions, (b)-(d) to the diamagnetic contributions. Here solid lines indicate the fermionic propagators, dashed lines refer to external photons, and solid circles denote current vertices.}
    \label{fig:FD1}
\end{figure*}

\begin{figure*}
    \centering
    \includegraphics[width=14cm]{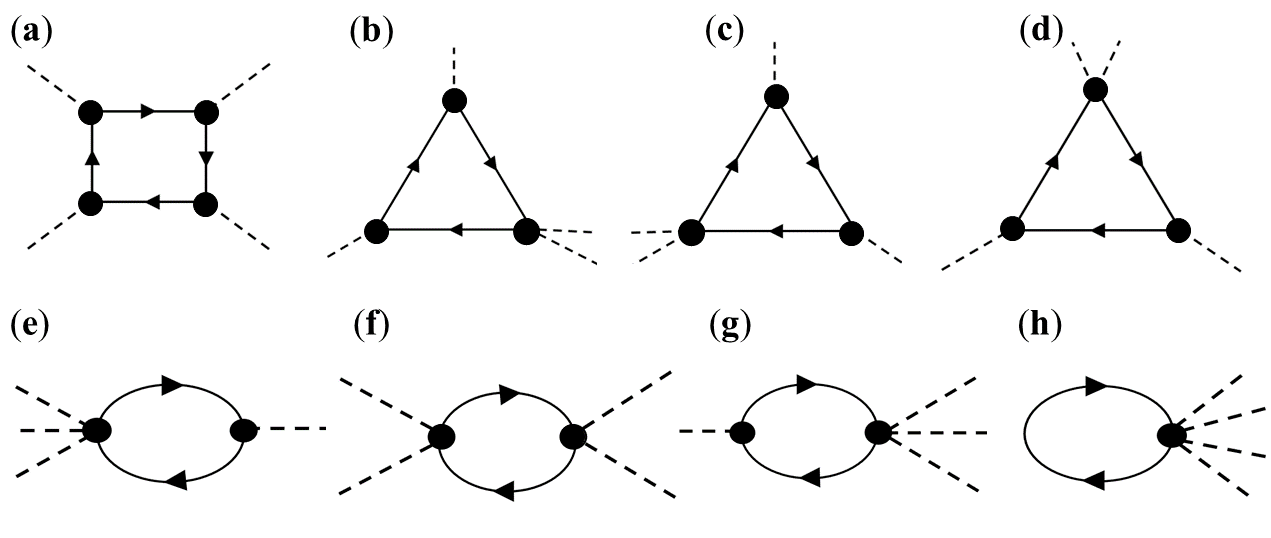}
    \caption{Feynman diagrams for the third-order response. (a) refer to the paramagnetic contributions, (b)-(h) to the diamagnetic contributions.}
    \label{fig:FD2}
\end{figure*}

To compute the nonlinear rectified spin Hall response of 1T$'$-WTe$_2$ to an external homogeneous vector potential ${\bm A}(t)$, we consider the corresponding driving electric field ${\bm E}(t) = -\partial_t {\bm A}(t)$. In a spin-diagonal basis, the nonlinear spin response is given as the difference of susceptibility for two spin polarisations:
\begin{align}
    \chi^{(n),{\rm spin}} =\chi^{(n)} (\uparrow)-\chi^{(n)} (\downarrow). 
\end{align}
The second-order rectified spin current thus follows
\begin{align} 
    \label{eqn:SOPG}
    j_a^{\rm PG} &= \sum_{b c} \frac{\chi_{abc}^{(2),\rm{spin}}(\omega,-\omega)} {(i\omega)(-i\omega)} E_b(\omega) E_c^*(\omega).
\end{align}
Similarly, the two-color third-order rectified spin current reads
\begin{align} 
    j_a^{\rm 2c-PG} &= \sum_{b c d}  \frac{\chi_{abcd}^{(3),\rm{spin}}(\omega,\omega,-2\omega)}{(i\omega)(i\omega) (-i2\omega)} E_b(\omega) E_c(\omega) E_d^*(2\omega),
    \label{eqn:DC}
\end{align}
where $\chi_{abc}^{(2),{\rm spin}}(\omega_1,\omega_2)$ and $\chi_{abcd}^{(3),{\rm spin}}(\omega_1,\omega_2,\omega_3)$ respectively stand for the second- and third-order spin-resolved susceptibility tensor elements in the two-dimensional $x-y$ coordinate. 
In addition, ${\bm E}(\omega)$ and ${\bm E}(2\omega)$ stand for the field elements at two colors with frequency $\omega$ and $2\omega$. We employ the diagrammatic perturbative approach, and the response functions are represented by the Feynman diagrams in Fig.~\ref{fig:FD1} and \ref{fig:FD2}. 

The second-order spin-resolved current given by the correlation of three paramagnetic current operators (or one-photon current coupling) $\hat j_a =(-e/\hbar) \partial_{k_a} \hat {\cal H}$ as $\chi^{(2),P}_{abc}\sim \langle \hat j_a \hat j_b \hat j_c\rangle $ and can be diagrammatically depicted as in Fig.~\ref{fig:FD1}(a) which formally reads 
\begin{align} \label{eqn:chi_para} 
    \chi_{abc}^{(2),P}(\omega_1,\omega_2,s) &= \sum_{\mathcal{P}} \sum_{\{\lambda_i \},\bm{k}} \frac{{j}_{a,{\bm k},s}^{\lambda_1\lambda_2}~{j}_{b,{\bm k},s}^{\lambda_2\lambda_3}~{j}_{c,{\bm k},s}^{\lambda_3\lambda_1}}{\hbar\omega_\Sigma + \varepsilon_{\bm{k},s}^{\lambda_2\lambda_1}}
    \left\{ 
    \frac{f^{\lambda_2 \lambda_3}_{{\bm k},s}}{\hbar\omega_1 + \varepsilon_{{\bm k},s}^{\lambda_2\lambda_3}}
    - 
    \frac{f^{\lambda_3 \lambda_1}_{{\bm k},s}}{\hbar\omega_2 + \varepsilon_{{\bm k},s}^{\lambda_3\lambda_1}}\right\},
\end{align}
where $\omega_{\Sigma} = \omega_1 + \omega_2$, $\varepsilon_{\bm{k},s}^{\lambda_i\lambda_j}= \varepsilon_{\bm{k},s}^{\lambda_i} - \varepsilon_{\bm{k},s}^{\lambda_j}$ is the energy difference between different bands with given spin index $s=\pm1$ ($+1$ for up-spin and $-1$ for down-spin) and $f^{\lambda_i\lambda_j}_{{\bm k},s} = f(\varepsilon_{\bm{k},s}^{\lambda_i})-f(\varepsilon_{\bm{k},s}^{\lambda_j})$ refers to the difference between the Fermi function corresponding to distinct bands. Here $f(\varepsilon) = [1+e^{\beta(\varepsilon-\mu)}]^{-1}$ is the Fermi-Dirac distribution function, $\mu$ is the chemical potential, and $\beta = 1/k_B T$ having $k_B$ as the Boltzmann constant, $T$ as an electron temperature.
The one-photon coupling vertex is ${j}_{a,{\bm k},s}^{\lambda_i \lambda_j} =  \langle u_{{\bm k},s}^{\lambda_i}| \hat j_{a} |u_{{\bm k},s}^{\lambda_j}\rangle$ with $|u_{{\bm k},s}^{\lambda_i}\rangle$ being an eigen vector of the Hamiltonian.   
Note that $ \sum_{\mathcal{P}}$ stands for the intrinsic permutation symmetry $(b,\omega_1) \Longleftrightarrow (c,\omega_2)$ and $\hbar\omega_i \rightarrow \hbar\omega_i + i\eta$ with the parameter $\eta \rightarrow 0^+$.
The diamagnetic contribution to the second-order response (Fig.~\ref{fig:FD1}(b)-(d)) is given by 
\begin{align} \label{eqn:chi_dia} \nonumber
    \chi_{abc}^{(2),D}(\omega_1,\omega_2,s) &= -\sum_{\mathcal{P}} \sum_{\{\lambda_i \},\bm{k}} \bigg\{ {\xi}_{abc,{\bm k},s}^{\lambda_1\lambda_1} f(\varepsilon_{\bm{k},s}^{\lambda_1})  \nonumber\\
    & 
    + {j}_{a,{\bm k},s}^{\lambda_1\lambda_2}~{\kappa}_{bc,{\bm k},s}^{\lambda_2\lambda_1}
    \frac{f^{\lambda_1 \lambda_2}_{{\bm k},s}}{\hbar\omega_1 + \varepsilon_{{\bm k},s}^{\lambda_1\lambda_2}}
     +
      \frac{f^{\lambda_1 \lambda_2}_{{\bm k},s}}{2}
      \frac{{j}_{b,{\bm k},s}^{\lambda_2\lambda_1}~{\kappa}_{ac,{\bm k},s}^{\lambda_1\lambda_2}+{j}_{c,{\bm k},s}^{\lambda_2\lambda_1}~{\kappa}_{ba,{\bm k},s}^{\lambda_1\lambda_2}}{\hbar\omega_\Sigma + \varepsilon_{{\bm k},s}^{\lambda_1\lambda_2}}  
     \bigg\}.
\end{align}
Notice that the two-photon and three-photon current couplings are $\hat{\kappa}_{ab} = (-e/\hbar)^2 \partial_{k_a}\partial_{k_b} \hat{\mathcal{H}}$ and $\hat\xi_{abc} = - (1/2) (e/\hbar)^3 \partial_{k_a} \partial_{k_b} \partial_{k_c} \hat {\cal H}$, respectively. 
It is important to emphasize that our diagrammatic method is highly systematic, allowing us to include all contributions, including both paramagnetic and diamagnetic terms, to second and third-order currents. Consequently, the final result adheres to gauge-invariant sum-rules~\cite{rostami_AP_2021}. Furthermore, our low-energy model Hamiltonian approach produces physically meaningful results. This is due to the rapid convergence of wave-vector integration for large ${\bm k}$ values, making our final result independent of the choice of the ultraviolet cutoff.

Similarly, following the diagrammatic scheme in Fig.~\ref{fig:FD2} (a), the third-order spin response originating from the paramagnetic current operator is formally given by $\chi^{(3),P}_{abcd}\sim \langle \hat j_a \hat j_b \hat j_c \hat j_d \rangle$.  In principle, there are seven distinct Feynman diagrams involving multi-photon current couplings~\cite{rostami_AP_2021}. 
However, for our Hamiltonian model, only four additional diagrams can contribute, as depicted in Fig.~\ref{fig:FD2}(b,c,d,f), which can be written in terms of the following correlation functions $\langle\hat \kappa_{ab}\hat \kappa_{cd}\rangle$, $\langle \hat j_a \hat j_b \hat \kappa_{cd}\rangle $, $\langle \hat \kappa_{ab} \hat j_c \hat j_d\rangle$ and  $\langle \hat j_a \hat \kappa_{bc}  \hat j_d\rangle$. We provide the explicit form of the third-order response in~\ref{app:chi3} for brevity.

\subsection{Nonlinear spin Hall current}
The $n^{\rm th}$-order nonlinear response functions to external fields are dictated by the susceptibility tensor components $\chi_{abc\cdots}^{(n)}$. The tensor components are restricted by space inversion, time-reversal, and rotational and mirror symmetries of the crystalline quantum material. Specifically, for a system having mirror symmetry $\mathcal{M}_a$ along the plane perpendicular to an arbitrary spatial axis $a$ and the spin polarization direction, the susceptibility tensor components with an odd number of `$a$' spatial indices do not contribute to the charge current. This arises due to the cancellation of the up and down spin components as $\chi^{(n)} = \chi^{(n)}_{\uparrow} + \chi^{(n)}_{\downarrow} = 0$. But, these tensor components contribute to the spin response function $\chi^{(n), {\rm spin}} = \chi^{(n)}_{\uparrow} - \chi^{(n)}_{\downarrow}=2 \chi^{(n)}_{\uparrow}$. 

In the time-reversal symmetric WTe$_2$ with broken inversion symmetry due to vertical gate potential, the mirror symmetry with respect to $x$-axis, i.e., $\mathcal{M}_x$ reduces non-vanishing tensor elements to four and eight for the second-order and third-order responses respectively. 
These second-order components are 
$\chi_{xyy}^{(2)}, \chi_{xxx}^{(2)},  \chi_{yyx}^{(2)}, \chi_{yxy}^{(2)}$. 
For the third-order response, the non-vanishing tensor elements are the following 
\begin{align}    
    &\chi_{yxxx}^{(3)}, \chi_{xyyy}^{(3)}, \chi_{xxxy}^{(3)}, \chi_{xyxx}^{(3)},
    \nonumber \\
    &\chi_{xxyx}^{(3)}, \chi_{yxyy}^{(3)}, \chi_{yyxy}^{(3)}, \chi_{yyyx}^{(3)}.
\end{align}
To demonstrate the polarization dependence, we decompose the nonlinear spin Hall current in the longitudinal and transverse basis as ${\bm j}^{\rm spin} = j^{\rm spin}_{||} \hat {\bm \epsilon}_{||} + j^{\rm spin}_{\perp} \hat {\bm \epsilon}_\perp$. Here, the unit vectors are orthogonal $\hat{\bm \epsilon}_{||}\cdot\hat {\bm \epsilon}_{\perp}=0$.

Accordingly, the longitudinal (parallel) and transverse (perpendicular) components of the spin current can be formulated using the relation $j^{\rm spin}_{||} = {\bm j}^{\rm spin}\cdot \hat{\bm \epsilon}_{||}$ and $j^{\rm spin}_{\perp} = {\bm j}^{\rm spin}\cdot \hat {\bm \epsilon}_{\perp}$ respectively. 
In this study, we focus on the transverse (Hall) components of the nonlinear rectified spin current due to single-color driving field ${\bm E}(t)= |E_1|e^{i\phi_1} \hat{\bm \epsilon}(\theta) e^{i\omega t}+c.c.$ and two-color light field ${\bm E}(t)= \hat{\bm \epsilon}(\theta) \left [|E_1|e^{i(\omega t+\phi_1)} + |E_2| e^{i(2\omega t + \phi_2)} + c.c.\right]$ where $\hat {\bm \epsilon}_{||}=\hat{\bm \epsilon}(\theta)=\cos\theta \hat{\bm x} + \sin\theta \hat{\bm y}$, with $\theta$ being the polarization angle, is the linear polarization unit vector. 
The corresponding light-induced single-color and two-color nonlinear spin-current are given by $j_{\perp}^{\rm PG}$ and 
$j_{\perp}^{\rm 2c-PG}$ respectively. 
Symmetry-based argument (as discussed earlier) leads to the following polarisation dependence of the transverse component of the nonlinear spin currents:
\begin{align}
\label{eqn:SOjPG}
    &j_{\perp}^{\rm PG} =  \frac{|E_1|^2}{\omega^2} {\rm Re}\left[\sin^2\theta \chi_{xyy}^{(2)} + \cos^2\theta \chi_{xxx}^{(2)}\right]\sin\theta,
\end{align}
and the third-order spin Hall current is given by 
\begin{align}
    &j_{\perp}^{\rm 2c-PG} =  \frac{|E_1|^2 |E_2|}{\omega^3}{\rm Im} \Big[  e^{i\Delta \phi}\Big(\chi^{(3)}_1  
    \sin^4\theta -\chi^{(3)}_2 %
    \cos^4\theta 
    + \chi^{(3)}_3 \sin^2 2\theta \Big) \Big],
   \label{eqn:TOHC}
\end{align}
where $\chi^{(3)}_1=\chi_{xyyy}^{(3)}$, $\chi^{(3)}_2=\chi_{yxxx}^{(3)}$, and
\begin{align}
\chi^{(3)}_3=\frac{\chi_{xxxy}^{(3)} + \chi_{xxyx}^{(3)} + \chi_{xyxx}^{(3)} - \chi_{yxyy}^{(3)} - \chi_{yyxy}^{(3)} -  \chi_{yyyx}^{(3)}}{4}. 
\end{align}
The derivation of the nonlinear spin currents is provided in ~\ref{app:App3}.

Further, $\Delta\phi = (2\phi_1-\phi_2)$ is the phase difference between the single- and second-harmonic frequency beams. Here, we get two sets of terms that are proportional to $ {\rm Re}[\chi^{(3)}]\sin\Delta\phi$ and  ${\rm Im}[\chi^{(3)}] \cos\Delta\phi$ contribute to the third-order rectified spin Hall current depending on the choice of phase difference. It has been found that for the linearly polarized beams with $\Delta\phi = \pm \pi/2$, the nonlinear Hall current Eq.~\eqref{eqn:TOHC} yields maximum value as discussed in Ref.~\cite{rioux_PRB2011} for the case of graphene. In this study, we consider zero phase difference $\Delta\phi=0$.

The following section presents numerical results in gated single-layer 1T$'$-WTe$_2$ for the second and third-order spin susceptibilities versus laser frequency and the gate potential $U$. Afterward, we analyze the anisotropic polarisation dependence of the nonlinear spin Hall susceptibilities.

\section{Numerical Results and Discussion}

\begin{figure*}[htbp]
    \centering
    \includegraphics[width=17cm]{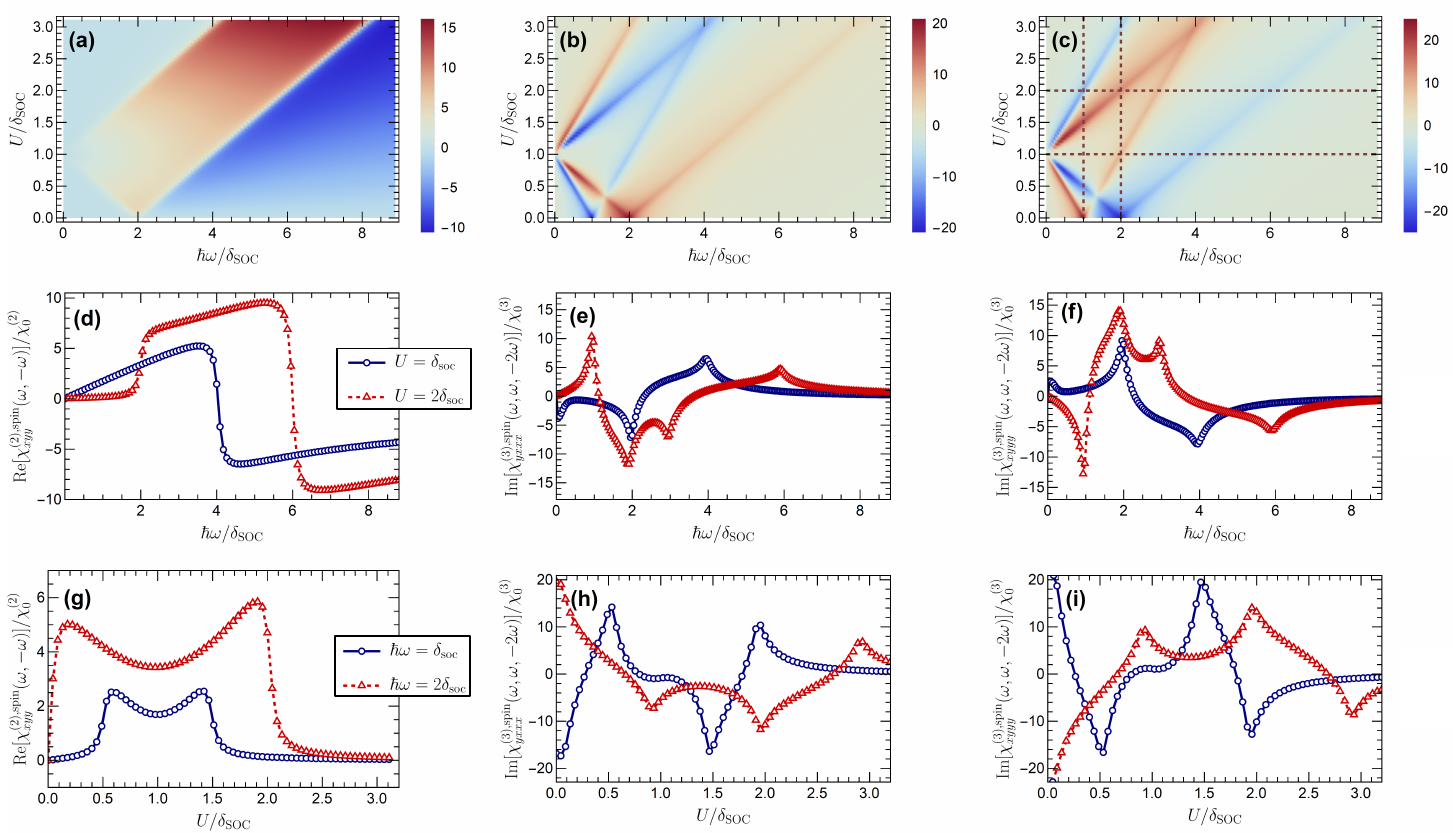}
    \caption{Top: (a)-(c) Density plots for the second-order and third-order spin Hall response as a function of the normalized incident energy and normalized out-of-plane gate potential with respect to the spin-orbit coupling gap. The dotted lines in plot (c) refer to the values for the cross-section plots in the second and third-row panels, and the color bar refers to the magnitude of the spin response. Second row: (d)-(f) represent the rectified spin Hall response $\chi_{xyy}^{(2)}$, $\chi_{yxxx}^{(3)}$ and $\chi_{xyyy}^{(3)}$ at out-of-plane gate potential $U=\delta_{\rm{SOC}}$ and $U=2\delta_{\rm{SOC}}$. Third row: (g)-(i) represent the response at an incident energy $\hbar\omega=\delta_{\rm{SOC}}$ and $\hbar\omega=2\delta_{\rm{SOC}}$. Here, we set the chemical potential $\mu=0$ eV and set other parameters based on the DFT calculations~\cite{qian_Sc2014} as $v_x =0.09$ eV\AA, $v_y = 2.6$ eV\AA, $\delta = -0.33$ eV, $B = 5.28$ eV\AA$^2$, $A = -2.64$ eV\AA$^2$, and $T=1$ K. Further, we normalize the response by setting $\chi_0^{(2)} = e^3/\hbar$ and $\chi_0^{(3)} = e^4/\hbar$.}
    \label{fig:Fig4}
\end{figure*}

We find that there is no second-order spin current ($\propto \chi^{(2)}_{yxx}$) in the Hall measurement geometry along $\hat{y}$-direction with the field along $\hat{x}$-direction due to the cancellation of the current induced by up and down spins, but the second-order Hall charge current will remain finite. 
Unlike the second-order current, the third-order spin Hall current $j_a^{(3),\text{spin}}$ follows different symmetry properties, thus can be obtained in both $\hat{x}$ and $\hat{y}$ Hall measurement geometries. Specifically, the third-order spin Hall effect can be observed in both inversion symmetric and time-reversal symmetric materials. Here, we discuss the two Hall components of susceptibility $\chi_{yxxx}^{(3),\text{spin}}$, and $\chi_{xyyy}^{(3),\text{spin}}$, subjected to the mirror symmetry possessed by 1T$'$-WTe$_2$, to the linearly polarized two-color optical beam of light. 

In Fig.~\ref{fig:Fig4} (a)-(c), we demonstrate the density plot for the second- and third-order spin Hall susceptibility tensor components $\chi_{xyy}^{(2),{\rm{spin}}}$, $\chi_{yxxx}^{(3),{\rm{spin}}}$ and $\chi_{xyyy}^{(3),{\rm{spin}}}$ for 1T$'$-WTe$_2$ having spin texture in the momentum space, as a function of $U/\delta_{\rm{SOC}}$ and $\hbar\omega/\delta_{\rm{SOC}}$, where $\delta_{\rm SOC}$ is the spin-orbit coupling gap. Here, we mainly focus on the rectified (dc) spin current with single-color and two-color laser beams, contributed by the real part of the second-order and imaginary part of the third-order response functions, respectively. 
We observe that the second-order spin Hall susceptibility component $\chi_{xyy}^{(2),{\rm spin}}$ in response to the electric field along $\hat{y}$ direction gives interband resonances at an incident energy $\hbar\omega \ge 2\Delta_{\pm}$, with $\Delta_{\pm}= 2|U \pm \delta_{\rm{SOC}}|$ being the band gap at $Q$ and $Q'$, as indicated by the dark red and dark blue colors in Fig.~\ref{fig:Fig4}(a). However, resonances' strength and position change with the out-of-plane gate potential, thus shifting in different frequency regimes. 

The resulting feature is intriguing because of the band closing and reopening around the Dirac points at valleys $Q$ and $Q'=-Q$ due to the gate potential $U$. Furthermore, the spin current generation occurs due to the imbalance of carrier excitation at different valleys.
Since spin-up and spin-down currents are equal in magnitude and flow in opposite directions, the charge Hall current vanishes and we obtain a pure second-order spin Hall current by irradiating the linearly polarized single-color beam of light. 
For the case of zero phase difference, the two-color rectified spin current is given by the real part of the third-order conductivity (see~\ref{app:App3}) 
\begin{align}
    j_\perp^{\rm 2c-PG} &= {\rm Re}\left[ \hat {\bm \epsilon}_\perp \cdot {\bm \sigma}^{(3)}(\omega,\omega,-2\omega) \vdots \hat {\bm \epsilon}_{||}\hat {\bm \epsilon}_{||}\hat {\bm \epsilon}_{||} \right] |E(\omega)|^2|E(2\omega)|.
\end{align}

One of the most striking results of this study is the presence of light-induced nonlinear spin current when $\hbar\omega < \Delta_{\pm}$. Unlike usual shift and injection current mechanisms which are related to the imaginary part ${\rm Im}[1/(\Delta_{\pm}-\hbar\omega -i0^+)]$, the current below the optical gap is given by the real part ${\rm Re}[1/(\Delta_{\pm}-\hbar\omega -i0^+)]$. In the clean limit, the imaginary part leads to a delta function $\delta(\Delta_{\pm}-\hbar\omega)$ implying that we can induce a current after absorbing the photon and generating a photo-excited carrier. However, the current below the optical gap is related to the principal value of $1/(\Delta_{\pm}-\hbar\omega -i0^+)$ and thus can be finite even without generating a photo-excited carrier density. This mechanism has been discussed for the second-order charge current in the time-reversal symmetry breaking systems~\cite{kalpan_PRL2020,gao_PRR2021}, and its thermodynamics ground is discussed in detail~\cite{shi_PRB2023}. 

In Ref.~\cite{kalpan_PRL2020}, the second-order charge conductivity below the optical gap regime has been predicted for the time-reversal symmetry broken system. The conductivity is given by the expression $\sigma^{(2)}_{aaa}\sim \sum_{m \neq n} \sum_{{\bm k},s} f^{mn}_{{\bm k},s}|r^{mn}_{a,{\bm k},s}|^2 v^{m}_{a,{\bm k},s}$, where $f^{mn}_{{\bm k},s}=f(\varepsilon^{m}_{{\bm k},s})-f(\varepsilon^{n}_{{\bm k},s})$. In this expression, $r^{mn}_{a,{\bm k},s}$ represents the inter-band matrix element of the position vector Cartesian component labeled by the subscript $a$, and $v^{m}_{a,{\bm k},s}= \hbar^{-1} \partial_{k_a} \varepsilon^{m}_{{\bm k},s}$ denotes the band velocity component. It is worth noting that the current vanishes in the presence of time-reversal symmetry, as the velocity is odd under time-reversal.

Our numerical results in Fig.~\ref{fig:Fig4} show that we can extend the above result of Ref. \cite{kalpan_PRL2020} to the third-order regime in the spin channel without breaking the time-reversal symmetry, offering an intriguing extension to this exotic behavior.
To shed further light, we also provide the explicit analysis in~\ref{app:App4}. Here, we present an intuitive sketch of the analytical expression for the third-order spin current within the optical gap regime. The concept is based on replacing the band velocity with the anomalous velocity $(-e/\hbar){\bm E} \times {\bm \Omega}^m_{{\bm k}}$, which allows us to obtain the third-order charge current. In this case, ${\bm \Omega}^m_{\bm k}={\bm \Omega}^m_{{\bm k}\uparrow}+{\bm \Omega}^m_{{\bm k}\downarrow}$ represents the Berry curvature vector.
To obtain the spin current, we can roughly replace the Berry curvature with the spin-Berry curvature ${\bm \Omega}_{{\bm k}}^{m,{\rm spin}}={\bm \Omega}^{m}_{{\bm k},\uparrow}-{\bm \Omega}^{m}_{{\bm k},\downarrow}$. Consequently, the third-order spin current can be finite in the presence of time-reversal symmetry. The explicit calculation given in~\ref{app:App4} is slightly different but qualitatively consistent with the above intuitive picture. 
Below the gap, the third-order effect is given by the spin Hall current $ j^{\rm 2c-PG}_{y}= \sigma^{\rm spin}_{yxxx} E^2_x(\omega) E^\ast_x(2\omega)$ where the nonlinear spin-Hall conductivity for inversion symmetric case ($U=0$) in our two-band model is given in terms of conduction band spin-Berry curvature:   
\begin{align}
 \sigma^{\rm spin}_{yxxx} \approx C(\omega) \sum_{\bm k} \Omega^{c,\rm spin}_{\bm k} \alpha^{cv}_{\bm k} 
\end{align}
where $C (\omega)= e^3/(2\hbar^4\omega^3)$ and the real-valued parameter $\alpha_{\bm k}^{cv}$ is given by 
\begin{align} \label{eqn:SpinCond} 
\alpha^{cv}_{\bm k} = \frac{1}{2} \sum_{s=\pm}
    \bigg\{|r^{cv}_{x,{{\bm k},s}}|^2 \varepsilon_{{\bm k},s}^{cv}  + 
     \tilde{\kappa}^{cv}_{xx,{\bm k},s} \bigg\}f^{cv}_{{\bm k},s}, 
\end{align}
in which $\tilde{\kappa}^{cv}_{xx,{\bm k},s} = (\hbar^2/e^2)\kappa^{cv}_{xx,{\bm k},s}$, and the position vector matrix element is given in terms of the paramagnetic current matrix element:
$r^{cv}_{a, {\bm k},s}  = ({i\hbar}/{e}) {{j}^{cv}_{a,{\bm k},s}}/{\varepsilon^{cv}_{{\bm k},s}}$
Considering time-reversal and inversion symmetries, we have $\Omega^{c}_{\bm k,\uparrow} = -\Omega^{c}_{\bm k,\downarrow}= -\Omega^{c}_{-\bm k,\downarrow}$, therefore $\Omega^{c}_{\bm k} = 0$, and $\Omega^{c,\mathrm{spin}}_{\bm k} = \Omega^{c,\mathrm{spin}}_{-\bm k}$. Similarly, we have $\varepsilon^{cv}_{\bm k,\uparrow} = \varepsilon^{cv}_{-\bm k,\downarrow}$, which leads to $\alpha^{cv}_{\bm k} = \alpha^{cv}_{-\bm k}$.
Accordingly, both $\alpha^{cv}_{\bm k}$ and spin-Berry curvature $\Omega^{c,\mathrm{spin}}_{\bm k}$ are even function under $\bm k \to - \bm k$. Consequently, the third-order spin Hall current below the optical gap is expected to be finite, which is consistent with our numerical results shown in Fig. \ref{fig:Fig4}.
%
%

In the following, we show that one can generate third-order spin current below the gap in a time-reversal symmetric system. However, there is no such effect in the second-order spin response due to the cancellation of the effect for spin-up and spin-down components.

To illustrate features with more detail, we plot the cross-section curves at different values of the scaled back-gate potential $U/\delta_{\rm{SOC}}$ and the incident photon energy $\hbar\omega/\delta_{\rm{SOC}}$. For $U = 0$, the inversion symmetry of the system remains intact, which results in the vanishing second-order response $\chi_{xyy}^{(2)}$  consistent with the symmetry argument. For finite but small $U\ll\delta_{\rm SOC}$, the inversion symmetry breaks down that leads to finite $\chi_{xyy}^{(2)}$.

\begin{figure*}
    \centering
    \includegraphics[width=15cm]{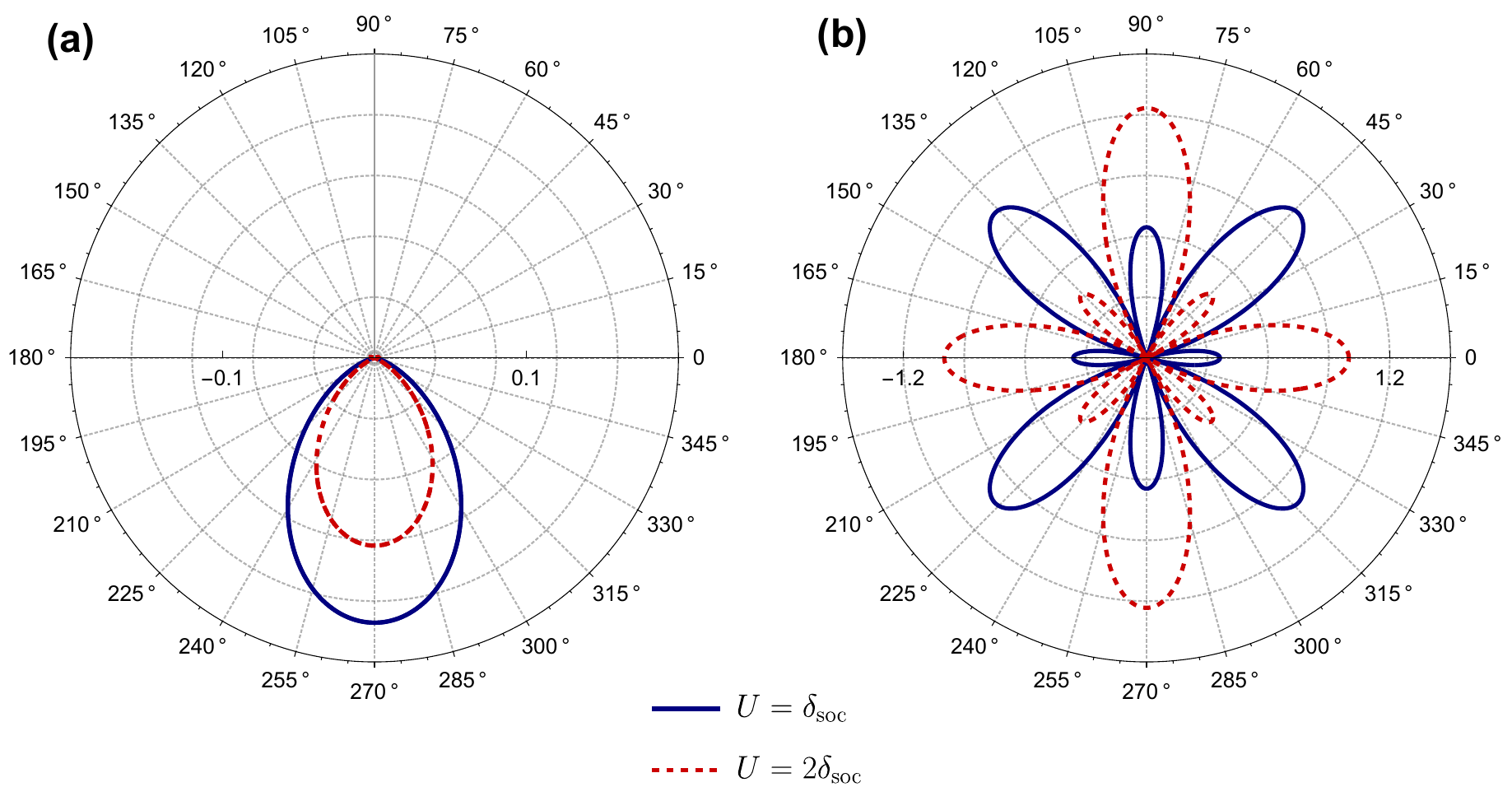}
    \caption{Polarization dependence of the second-order and third-order tangent component of the spin response at fixed normalized incident energy and different normalized out-of-plane gate potential with respect to the spin-orbit coupling gap. The red color curve correspond to $j_{\perp}^{{\rm PG, {\rm spin}}}/j_0^{(2)}$ and $0.1 j_{\perp}^{{\rm 2c-PG, {\rm spin}}}/j_0^{(3)}$ at $U=2\delta_{\rm SOC}$ and the blue refers to $0.1 j_\perp^{{\rm PG, {\rm spin}}}/j_0^{(2)}$ and $j_\perp^{{\rm 2c-PG, {\rm spin}}}/j_0^{(3)}$ at $U=\delta_{\rm SOC}$. }
    \label{fig:Fig5}
\end{figure*}

As seen in Fig.~\ref{fig:Fig4}(d), the effective band gap $\Delta_-$ vanishes, and the bottom of the conduction and top of the valence bands touch each other at $Q$ point when $U=\delta_{\rm SOC}$ and therefore optical response reveals the gapless nature of the system that is finite even at the tiny value of the light frequency. By further increasing the frequency, we see a step-like feature in the second-order response at $\hbar\omega=2\Delta_{+}=4U$. We have generated the same curve for the case of $U=2\delta_{\rm SOC}$ where the effective gaps at different valleys are $\Delta_+/\delta_{\rm{SOC}} = 3$ and $\Delta_-/\delta_{\rm{SOC}} = 1$. This leads to the resonant jumps at $\hbar\omega/\delta_{\rm{SOC}} = 6$ and $\hbar\omega/\delta_{\rm{SOC}} = 2$.

In Fig.~\ref{fig:Fig4}(e)-(f), we find the emergence of resonant peaks in the third-order response due to the two-photon and one-photon absorption processes which correspond to $\hbar\omega=\Delta_{\pm}$ and $\hbar\omega=2\Delta_{\pm}$, respectively. Among these, the two peaks associated with a larger effective gap $\Delta_+$ show similar behavior, such as resonances for different back-gate potentials. 
However, the other two associated with $\Delta_-$ reveal the different feature that is the presence and absence of the resonance peak at $U=\delta_{\rm{SOC}}$ and $U=2\delta_{\rm{SOC}}$. 

We have made vertical cross-section plots to investigate the behavior of nonlinear spin susceptibilities concerning fixed values of photon energy $\hbar\omega$ and varying values of $U$. The second-order and third-order effects are depicted in Fig.~\ref{fig:Fig4}(g) and Fig.~\ref{fig:Fig4}(h)-(i), respectively. Contrary to the second-order case, the third-order response remains finite at $U=0$. Furthermore, we observe a strong and non-monotonic dependence of the nonlinear spin current on the back-gate potential $U$. This behavior implies that small changes in the back-gate potential can induce significant changes in the resulting nonlinear spin response, which could have important implications for the design and optimization of spin-based devices.

Further, we present the numerical results for the polarisation dependence of the spin transverse currents for a fixed value of $\hbar\omega = \delta_{\rm SOC}$ in polar plots in Fig.~\ref{fig:Fig5}. The polarization dependence of nonlinear spin current highly depends on the crystalline symmetry and orientation of the monolayer WTe$_2$. The anisotropic behavior of the second-order spin Hall current that follows $j^{{\rm PG}}_{\perp}\propto {\rm Re}[\chi^{(2)}_{xyy}] \sin^3\theta $. The third-order spin current is proportional to the imaginary part of $\chi^{(3)}$ components for zero phase difference between two color laser pulses. The anisotropic polarization dependence of third-order spin current is more complex due to the competing behavior of different tensor components of the susceptibility multiplied by $\sin^4\theta$, $\cos^4\theta$, and $\sin^22\theta$.

Finally, we provide the numerical estimation of the spin Hall photocurrents at a particular incident energy and using the microscopic parameter obtained from DFT calculations and experiments~\cite{xu_NP2018, garcia_PRL2020,tang_NP2017}. At $\hbar\omega = 90$ meV, which is twice the spin-orbit coupling gap $\delta_{\rm{SOC}} = 45$ meV, the out-of-plane gate potential $U = 90$ meV, and the electric field $E$ of the order of $(10^{6} - 10^{9})$ V/m, the magnitude of the second and third-order rectified spin Hall conductivity comes out to be of the order of $(10^{-1}-10^{2})\sigma_0$ and $(10^{-4}-10^2)\sigma_0$ respectively, where $\sigma_0 = 2e/h$ is the unit of spin conductivity. Here, the back-gate potential $U$ is obtained using the relation $U \approx e\epsilon E_z d$ and setting vertical field $E_z \approx 0.1$V/nm, dielectric constant $\epsilon \approx 3$, and the separation between the top and bottom contacts $d \approx 0.31$ nm \cite{akash_NPJ2018, shah_PRB2019}. Compared to the linear dc spin conductivity, the presented rectified nonlinear spin Hall conductivity estimation is feasible to measure experimentally. 


Experimentally, the observation of nonlinear spin currents can be achieved through two distinct techniques: the inverse-spin Hall setup and magneto-optical Kerr rotation measurements. In the inverse-spin Hall setup, the flow of the spin current can be observed by measuring the transverse voltage drop (charge current flow) due to the inverse spin Hall effect~\cite{saitoh_APL2006, ando_JAP2011, bottegoni_APl2013}. The alternative approach involves magneto-optical spectroscopy of spin current, employing spatially-resolved Kerr rotation measurements-- as utilized to observe the spin-Hall effect in a 2D electron liquid~\cite{kato_Sc2004}. Light-induced nonlinear spin current induces an electron spin accumulation at the edges of the sample. The spatial distribution of this spin accumulation (local magnetization) can be experimentally probed via the Kerr rotation spectroscopy.

\section{Conclusion}
Our study focuses on the investigation of the rectified nonlinear spin Hall current in the 1T$'$-WTe$_2$ material, utilizing both single-color and two-color laser beams. We have conducted an extensive analysis on the effects of the displacement field, light intensity, and polarization direction on the nonlinear spin Hall current to gain a comprehensive understanding of this phenomenon.
Our findings indicate that the nonlinear response exhibits interband resonances arising from one-photon and two-photon absorption processes. It is worth noting that the two-color rectified response was found to be significantly stronger in magnitude than the single-color response when the displacement field was twice the spin-orbit coupling gap.
Our study provides valuable insights into the intrinsic nonlinear spin Hall effect in the 1T$'$-WTe$_2$ material and offers a path toward designing advanced optoelectronic devices capable of operating in nonlinear regimes for spintronics applications. Moreover, the present study can be extended to bilayer, trilayer and in general to few-layered WTe$_2$ systems. In particular, the intriguing ferroelectric transition observed in few-layered WTe$_2$~\cite{Fei_Na2018, wang_NPJCM2019, xiao_NP2020, sharma_SA2019} can impact the nonlinear spin Hall current. In such systems, the ferroelectric polarization will possibility enhance the effect by spontaneously breaking the inversion symmetry. The quantitative analysis of the nonlinear spin Hall effect in bilayer WTe$_2$ is left as the subject of a future study.


\section*{Acknowledgment}
This work was supported by Nordita and the Swedish Research Council (VR Starting Grant No. 2018-04252). Nordita is partially supported by Nordforsk. PB acknowledges the computational facilities provided by SRM-AP and is financially supported by Science and Engineering Research Board-State University Research Excellence under project number SUR/2022/000289. 

%

%
\onecolumngrid
\newpage
\appendix 
\section{Third-order susceptibility expression}
\label{app:chi3}
In a similar spirit of the second-order response, the paramagnetic contribution to the third-order response corresponding to the Feynman diagram Fig.~\ref{fig:FD2}(a) is given by \cite{Ahn2022,Antti_nc_2017}
\begin{align} \nonumber
    \chi_{abcd}^{P,(3)} (\omega_1, \omega_2, \omega_3,s) &= \sum_{\mathcal{P}}\sum_{{\bm k}} \sum_{\lambda_i} \frac{{j}_{a,{\bm k},s}^{\lambda_4\lambda_3}~{j}_{b,{\bm k},s}^{\lambda_3\lambda_2}~{j}_{c,{\bm k},s}^{\lambda_2\lambda_1}~{j}_{d,{\bm k},s}^{\lambda_1\lambda_4}}{\hbar\omega_\Sigma + \varepsilon_{{\bm k},s}^{\lambda_1\lambda_4}} \nonumber \\
    &
     \times \bigg\{ \frac{1}{\hbar\omega_1 + \hbar\omega_2 + \varepsilon_{{\bm k},s}^{\lambda_1\lambda_3}} \bigg(  \frac{f^{\lambda_1 \lambda_2}_{{\bm k},s}}{\hbar\omega_1 + \varepsilon_{{\bm k},s}^{\lambda_1\lambda_2}} - \frac{f^{\lambda_2 \lambda_3}_{{\bm k},s}}{\hbar\omega_2 + \varepsilon_{{\bm k},s}^{\lambda_2\lambda_3}} \bigg) \nonumber \\
     &
     - \frac{1}{\hbar\omega_2 + \hbar\omega_3 + \varepsilon_{{\bm k},s}^{\lambda_2\lambda_4}} \bigg( \frac{f^{\lambda_2 \lambda_3}_{{\bm k},s}}{\hbar\omega_2 + \varepsilon_{{\bm k},s}^{\lambda_2\lambda_3}} -\frac{f^{\lambda_3 \lambda_4}_{{\bm k},s}}{\hbar\omega_3 + \varepsilon_{{\bm k},s}^{\lambda_3\lambda_4}}  \bigg) \bigg\},
\end{align}
where $\omega_\Sigma = \omega_1 + \omega_2 + \omega_3$ and the diamagnetic part of the third-order susceptibility associated with Figs.~\ref{fig:FD2}(b)-(h) is given by \cite{rostami_AP_2021,Antti_nc_2017}
\begin{align} \nonumber
    \chi_{abcd}^{D,(3)} (\omega_1, \omega_2, \omega_3,s) &= \sum_{\mathcal{P}}\sum_{{\bm k}} \sum_{\lambda_i} \bigg\{ \frac{{\kappa}_{ab,{\bm k},s}^{\lambda_1\lambda_2}~{\kappa}_{cd,{\bm k},s}^{\lambda_2\lambda_1} ~ f^{\lambda_1 \lambda_2}_{{\bm k},s}}{\hbar\omega_1 + \hbar\omega_2 +\varepsilon_{{\bm k},s}^{\lambda_1\lambda_2}} \nonumber \\
    &
    -
    \frac{{j}_{a,{\bm k},s}^{\lambda_3\lambda_2}~{j}_{b,{\bm k},s}^{\lambda_2\lambda_1}~{\kappa}_{cd,{\bm k},s}^{\lambda_1\lambda_3}}{\hbar\omega_1+\hbar\omega_3 + \varepsilon_{{\bm k},s}^{\lambda_1\lambda_3}} \bigg(  \frac{f^{\lambda_1 \lambda_2}_{{\bm k},s}}{\hbar\omega_1 + \varepsilon_{{\bm k},s}^{\lambda_1\lambda_2}} - \frac{f^{\lambda_2 \lambda_3}_{{\bm k},s}}{\hbar\omega_2 + \varepsilon_{{\bm k},s}^{\lambda_2\lambda_3}} \bigg) \nonumber \\
    &
    - 
     \frac{{\kappa}_{ab,{\bm k},s}^{\lambda_3\lambda_2}~{j}_{c,{\bm k},s}^{\lambda_2\lambda_1}~{j}_{d,{\bm k},s}^{\lambda_1\lambda_3}}{\hbar\omega_\Sigma + \varepsilon_{{\bm k},s}^{\lambda_1\lambda_3}} \bigg(  \frac{f^{\lambda_1 \lambda_2}_{{\bm k},s}}{\hbar\omega_1+\hbar\omega_2 + \varepsilon_{{\bm k},s}^{\lambda_1\lambda_2}} - \frac{f^{\lambda_2 \lambda_4}_{{\bm k},s}}{\hbar\omega_2 + \varepsilon_{{\bm k},s}^{\lambda_2\lambda_4}} \bigg)\bigg\}.
\end{align}

\section{Nonlinear rectified currents}
\label{app:App3}
\subsection{Second-order rectified current}
The second-order rectified current is obtained by setting $\omega_1 + \omega_2 = 0$. In response to a time-dependent external electric field ${\bm E}(t) = \sum_{\omega} {\bm E}_1(\omega)e^{i(\omega t +\phi_1)} + {\rm c.c.}$, the current can be expressed as follows
\begin{align}
    j_a^{\rm PG} &= \sum_{b,c} \sigma_{abc}^{(2)}(\omega, -\omega) E_{1,b}(\omega) E_{1,c}^*(\omega) + {\rm c.c.}
\end{align}
Here, the conductivity is $\sigma_{abc}^{(2)} = -\chi_{abc}^{(2)}/\omega^2$. Further, on considering the spatial indices $b=c$ we get
\begin{align} \nonumber
    j_a^{\rm PG} &= \sum_{b} \left(\sigma_{abb}^{(2)}(\omega,-\omega) E_{1,b}(\omega) E_{1,b}^*(\omega) + \sigma_{abb}^{(2)}(-\omega,\omega) E_{1,b}^*(\omega) E_{1,b}(\omega) \right)\\ 
    j_a^{\rm PG} &= \sum_{b} \left(\sigma_{abb}^{(2)}(\omega,-\omega) + \sigma_{abb}^{(2)}(-\omega,\omega)\right) E_{1,b}(\omega) E_{1,b}^*(\omega) .
\end{align}
Note that $E_b^*(\omega) = E_b(-\omega)$. 
Expressing the conductivity in real and imaginary parts as $\sigma_{abb}^{(2)} = {\rm Re}[\sigma_{abb}^{(2)}] + i{\rm Im}[\sigma_{abb}^{(2)}]$ and using the relation $\sigma_{abb}^{(2)}(\omega,-\omega) =[ \sigma_{abb}^{(2)}(-\omega,\omega)]^\ast$, we obtain
\begin{align}
\label{eqn:jdc2}
    j_a^{(2),{\rm PG}} &= \sum_{b} {\rm Re}[\sigma_{abb}^{(2)}(\omega,-\omega)]E_{1,b}(\omega) E_{1,b}^*(\omega) .
\end{align}

\subsection{Third-order rectified current}
The third-order rectified current is obtained by setting $\omega_1 + \omega_2 + \omega_3 = 0$. In response to a time-dependent external electric field ${\bm E}(t) = \sum_{\omega} [{\bm E}_1(\omega)e^{i(\omega t+\phi_1)} + {\bm E}_2(\omega)e^{i(2\omega t+\phi_2)}] + {\rm c.c.}$, the current can be expressed as follows
\begin{align}
    j_a^{\rm 2c-PG} &= \sum_{b,c,d} \sigma_{abcd}^{(3)}(\omega, \omega,-2\omega) E_{1,b}(\omega)E_{1,c}(\omega) E_{2,d}^*(2\omega) + {\rm c.c.}
\end{align}
Here, the conductivity is $\sigma_{abcd}^{(3)} = -i\chi_{abcd}^{(3)}/2\omega^3$. Further, on considering the spatial indices $b=c=d$ we get
\begin{align}
    j_a^{\rm 2c-PG} &= \sum_{b} \left(\sigma_{abbb}^{(2)}(\omega,\omega,-2\omega) E_{1,b}(\omega) E_{1,b}(\omega)E_{2,b}^*(2\omega) + \sigma_{abbb}^{(3)}(-\omega,-\omega,2\omega) E_{1,b}^*(\omega) E_{1,b}^*(\omega)E_{2,b}(2\omega) \right).
\end{align}
Considering real-valued field amplitudes, we obtain 
\begin{align}
    j_a^{\rm 2c-PG} &= \sum_{b} \left(\sigma_{abbb}^{(2)}(\omega,\omega,-2\omega) 
 e^{i(2\phi_1-\phi_2)} + \sigma_{abbb}^{(3)}(-\omega,-\omega,2\omega) e^{-i(2\phi_1-\phi_2)}
    \right) |E_{1,b}|^2|E_{2,b}|.
\end{align}
Writing the conductivity in real and imaginary parts as $\sigma_{abbb}^{(3)} = {\rm Re}[\sigma_{abbb}^{(3)}] + i {\rm Im}[\sigma_{abbb}^{(3)}]$ and using the relation $\sigma_{abbb}^{(3)}(\omega,\omega,-2\omega) = [\sigma_{abbb}^{(3)}(-\omega,-\omega,2\omega)]^\ast$, we obtain
\begin{align}
\label{eqn:jdc3}
    j_a^{\rm 2c-PG} &= \sum_{b} {\rm Re}[ \sigma_{abbb}^{(3)}(\omega,\omega,-2\omega)  e^{i(2\phi_0-\phi_1)}] |E_{1,b}|^2|E_{2,b}|.
\end{align}

\subsection{Transverse component of the nonlinear spin currents}
For the single-color driving field ${\bm E}(t)= |E_1|e^{i\phi_1} \hat{\bm \epsilon}(\theta) e^{i\omega t}+c.c.$, the transverse nonlinear rectified current is given by
\begin{equation}
    j^{\rm PG}_{\perp} = {\bm j}^{\rm PG}\cdot \hat {\bm \epsilon}_\perp,
\end{equation}
where the unit vector $\hat {\bm \epsilon}_\perp = -\sin\theta \hat{\bm x} + \cos\theta \hat{\bm y}$ and $\hat {\bm \epsilon}_{||} = \cos\theta \hat{\bm x} + \sin\theta \hat{\bm y}$ is the linear polarization unit vector having $\theta$ an angle of polarization. Using this property, the transverse current becomes
\begin{equation}
    j^{\rm PG}_{\perp} = - j_x^{\rm PG} \sin\theta +  j_y^{\rm PG} \cos\theta. 
\end{equation}
Using Eq.~(\ref{eqn:jdc2}) and expressing $\sigma_{abc}(\omega_1,\omega_2) = \chi_{abc}(\omega_1,\omega_2)/(i\omega_1 i\omega_2)$ and considering the symmetry-permitted components of the second-order response in gated single-layer WTe$_2$, we obtain
\begin{equation}
    j^{\rm PG}_{\perp} = \frac{1}{\omega^2} {\rm Re}\left[\chi_{xyy}^{(2)} E_{1,y}E_{1,y}^* + \chi_{xxx}^{(2)} E_{1,x}E_{1,x}^* \right]\sin\theta.
\end{equation}
On further simplifications, the rectified second-order current yields
\begin{equation}
    j^{\rm PG}_{\perp} =  \frac{|E_1|^2}{\omega^2} {\rm Re}\left[\sin^2\theta \chi_{xyy}^{(2)} + \cos^2\theta \chi_{xxx}^{(2)}\right] \sin\theta.
\end{equation}
This derivation leads to Eq.~(\ref{eqn:SOjPG}) in the main text. 
Similarly, for the two-color light field ${\bm E}(t)= \hat{\bm \epsilon}(\theta) \left [|E_1|e^{i(\omega t+\phi_1)} + |E_2| e^{i(2\omega t + \phi_2)} + c.c.\right]$ the third-order rectified transverse current following Eq.~(\ref{eqn:jdc3}) gives
\begin{align} \nonumber
    j^{\rm 2c-PG}_{\perp} = \frac{1}{\omega^3} {\rm Im}\bigg[ \big\{\chi_{xyyy}^{(3)}  e^{i\Delta\phi} E_{1,y}^2E_{2,y}^* + (\chi_{xyxx}^{(3)} +\chi_{xxyx}^{(3)}+\chi_{xxxy}^{(3)})   e^{i\Delta\phi} E_{1,x}^2E_{2,y}^*\big\}\sin\theta \\ 
      - \big\{\chi_{yxxx}^{(3)}  e^{i\Delta\phi} E_{1,x}^2E_{2,x}^* + (\chi_{yxyy}^{(3)} +\chi_{yyxy}^{(3)}+\chi_{yyyx}^{(3)})   e^{i\Delta\phi} E_{1,y}^2E_{2,x}^*\big\}\cos\theta \bigg].
\end{align}
On substituting the field, the above equation reduces to
\begin{align} \nonumber
    j^{\rm 2c-PG}_{\perp} & = \frac{1}{\omega^3} {\rm Im}\bigg[ \big\{\chi_{xyyy}^{(3)}  \sin^4\theta + (\chi_{xyxx}^{(3)} +\chi_{xxyx}^{(3)}+\chi_{xxxy}^{(3)})    \sin^2\theta \cos^2\theta \\ 
    & - \chi_{yxxx}^{(3)}  \cos^4\theta + (\chi_{yxyy}^{(3)} +\chi_{yyxy}^{(3)}+\chi_{yyyx}^{(3)})   \sin^2\theta \cos^2\theta \big\}e^{i\Delta\phi} |E_{1}|^2|E_{2}| \bigg],
\end{align}
\begin{equation}
     j^{\rm 2c-PG}_{\perp} = \frac{|E_{1}|^2|E_{2}|}{\omega^3} {\rm Im}\bigg[ \big\{\chi_{1}^{(3)}  \sin^4\theta - \chi_{3}^{(2)} \cos^4\theta  + \chi_{3}^{(3)}\sin^22\theta  \big\}e^{i\Delta\phi}  \bigg].
\end{equation}
 where the quantities $\Delta\phi=2\phi_1-\phi_2$, $\chi^{(3)}_1=\chi_{xyyy}^{(3)}$, $\chi^{(3)}_2=\chi_{yxxx}^{(3)}$, and
$\chi^{(3)}_3=\{\chi_{xxxy}^{(3)} + \chi_{xxyx}^{(3)} + \chi_{xyxx}^{(3)} - \chi_{yxyy}^{(3)} - \chi_{yyxy}^{(3)} -  \chi_{yyyx}^{(3)}\}/{4}$. This derives the Eq.~(\ref{eqn:TOHC}) of the main text.

\section{Proof of Eq. (13)}
\label{app:App4}
Here, we provide a detailed derivation for the third-order spin Hall current in the transparent (low-frequency) regime.
The third-order conductivity due to the two-color excitation field is related to the following third-order susceptibility 
\begin{equation} \label{eqn:CSR}
    \sigma^{(3)}(\omega,\omega,-2\omega) = -i\frac{\chi^{(3)}(\omega,\omega,-2\omega)}{2\omega^3}.
\end{equation}
Using the many-body perturbation theory, the susceptibility $\chi^{(3)}$ is expressed as the sum of paramagnetic and diamagnetic contributions to the susceptibilities. 
The paramagnetic part of the third-order susceptibility due to two-color field is
\begin{align} \nonumber
    \chi_{abcd}^{P,(3)} (\omega, \omega, -2\omega) &= \sum_{\mathcal{P}}\sum_{{\bm k},s} \sum_{\lambda_i} \frac{{j}_{a,{\bm k},s}^{\lambda_4\lambda_3}~{j}_{b,{\bm k},s}^{\lambda_3\lambda_2}~{j}_{c,{\bm k},s}^{\lambda_2\lambda_1}~{j}_{d,{\bm k},s}^{\lambda_1\lambda_4}}{\varepsilon_{{\bm k},s}^{\lambda_1\lambda_4}} \nonumber \\ 
    & \bigg\{ \frac{1}{2(\hbar\omega+i\eta) + \varepsilon_{{\bm k},s}^{\lambda_1\lambda_3}} \bigg(  \frac{f^{\lambda_1 \lambda_2}_{{\bm k},s}}{(\hbar\omega+i\eta) + \varepsilon_{{\bm k},s}^{\lambda_1\lambda_2}} - \frac{f^{\lambda_2 \lambda_3}_{{\bm k},s}}{(\hbar\omega+i\eta) + \varepsilon_{{\bm k},s}^{\lambda_2\lambda_3}} \bigg)\nonumber \\
     &
     - \frac{1}{-(\hbar\omega+i\eta) + \varepsilon_{{\bm k},s}^{\lambda_2\lambda_4}} \bigg( \frac{f^{\lambda_2 \lambda_3}_{{\bm k},s}}{(\hbar\omega+i\eta) + \varepsilon_{{\bm k},s}^{\lambda_2\lambda_3}} -\frac{f^{\lambda_3 \lambda_4}_{{\bm k},s}}{-2(\hbar\omega+i\eta) + \varepsilon_{{\bm k},s}^{\lambda_3\lambda_4}}  \bigg) \bigg\},
\end{align}
where $\eta \rightarrow 0^+$. 
We use the following identity to separate real and imaginary parts 
\begin{equation}
   \lim_{~~~\eta\to 0^+} \frac{1}{\hbar\omega + i \eta} = {\rm P} \bigg( \frac{1}{\hbar\omega} \bigg) - i\frac{\pi}{\hbar} \delta(\omega),  
\end{equation}
where ${\rm P}(\dots)$ stands for the principal value part. 
The imaginary part of the paramagnetic contribution of the third-order susceptibility below the gap for the two-band case gives
\begin{align} \nonumber
\label{eqn:chiP}
    {\rm Im}[\chi_{yxxx}^{P,(3)} (\omega, \omega, -2\omega)] &= \sum_{\mathcal{P}}\sum_{{\bm k},s}
    \bigg\{
    \frac{{\rm Im}[{j}_{y,{\bm k},s}^{cv}~{j}_{x,{\bm k},s}^{vc}]|{j}_{x,{\bm k},s}^{cv}|^2}{\varepsilon_{{\bm k},s}^{cv}} 
    \frac{f^{cv}_{{\bm k},s}}{\varepsilon_{{\bm k},s}^{cv}} \bigg[  {\rm P} \bigg( \frac{1}{\hbar\omega + \varepsilon_{{\bm k},s}^{cv}}\bigg) + {\rm P} \bigg( \frac{1}{2\hbar\omega + \varepsilon_{{\bm k},s}^{cv}}\bigg)\bigg]\\
    & -
    \frac{{\rm Im}[{j}_{y,{\bm k},s}^{vc}~{j}_{x,{\bm k},s}^{cv}]|{j}_{x,{\bm k},s}^{vc}|^2}{\varepsilon_{{\bm k},s}^{vc}} 
    \frac{f^{vc}_{{\bm k},s}}{\varepsilon_{{\bm k},s}^{vc}} \bigg[  {\rm P} \bigg( \frac{1}{\hbar\omega - \varepsilon_{{\bm k},s}^{cv}}\bigg) + {\rm P} \bigg( \frac{1}{2\hbar\omega - \varepsilon_{{\bm k},s}^{cv}}\bigg)\bigg]
    \bigg\}.
\end{align}
On further simplifications and using the relations ${\rm Im}[{j}_{y,{\bm k},s}^{cv}{j}_{x,{\bm k},s}^{vc}] = - {\rm Im}[{j}_{y,{\bm k},s}^{vc}{j}_{x,{\bm k},s}^{cv}]$, $f^{cv}_{{\bm k},s} = -f^{vc}_{{\bm k},s}=f^{c}_{{\bm k},s}-f^{v}_{{\bm k},s}$ and $\varepsilon_{{\bm k},s}^{cv} = -\varepsilon_{{\bm k},s}^{vc}=\varepsilon_{{\bm k},s}^{c}-\varepsilon_{{\bm k},s}^{v}$, Eq.~\eqref{eqn:chiP} reduces to
\begin{align} \nonumber
    {\rm Im}[\chi_{yxxx}^{P,(3)} (\omega, \omega, -2\omega)] &= \sum_{{\bm k},s}
    \frac{{\rm Im}[{j}_{y,{\bm k},s}^{cv}~{j}_{x,{\bm k},s}^{vc}]|{j}_{x,{\bm k},s}^{cv}|^2}{\varepsilon_{{\bm k},s}^{cv}} \frac{f^{cv}_{{\bm k},s}}{\varepsilon_{{\bm k},s}^{cv}}
    \bigg\{
     \bigg[  {\rm P} \bigg( \frac{1}{\hbar\omega + \varepsilon_{{\bm k},s}^{cv}}\bigg) + {\rm P} \bigg( \frac{1}{2\hbar\omega + \varepsilon_{{\bm k},s}^{cv}}\bigg)\bigg]\\
    & -
    \bigg[  {\rm P} \bigg( \frac{1}{\hbar\omega - \varepsilon_{{\bm k},s}^{cv}}\bigg) + {\rm P} \bigg( \frac{1}{2\hbar\omega - \varepsilon_{{\bm k},s}^{cv}}\bigg)\bigg]
    \bigg\}.
\end{align}
In the regime $\hbar\omega \ll \Delta$, the third-order susceptibility yields
\begin{align} 
    &{\rm Im}[\chi_{yxxx}^{P,(3)} (\omega, \omega, -2\omega)] \approx \sum_{{\bm k},s}
    \frac{{\rm Im}[{j}_{y,{\bm k},s}^{cv}~{j}_{x,{\bm k},s}^{vc}]|{j}_{x,{\bm k},s}^{cv}|^2}{(\varepsilon_{{\bm k},s}^{cv})^2} \frac{f^{cv}_{{\bm k},s}}{\varepsilon_{{\bm k},s}^{cv}}.
\end{align}
Utilizing the definitions of the Berry curvature in the conduction band~\cite{xiao_RMP2010}, denoted as ${\Omega}^{c}_{{\bm k},s}$, and the valence band, denoted as ${\Omega}^{v}_{{\bm k},s}$, we can write 
\begin{align}
{\Omega}^{c}_{{\bm k},s} =-{\Omega}^{v}_{{\bm k},s} = 2 \left(\frac{\hbar}{e}\right)^2 {\rm Im}\frac{[{j}_{y,{\bm k},s}^{cv}~{j}_{x,{\bm k},s}^{vc}]}{(\varepsilon_{{\bm k},s}^{cv})^2},     
\end{align}
and the current matrix element in terms of the position vector matrix element reads
\begin{align}
 r^{cv}_{a,{\bm k},s} = \frac{i\hbar}{e} \frac{j^{cv}_{a,{\bm k},s}}{\varepsilon^{cv}_{{\bm k},s}}.  
\end{align}
Therefore, the third-order spin Hall susceptibility becomes 
\begin{align} \nonumber
    &{\rm Im}[\chi_{yxxx}^{P,(3)} (\omega, \omega, -2\omega)] \approx \frac{e^4}{2\hbar^4}  
     \sum_{{\bm k}}
    {\Omega}^{c,{\rm spin}}_{{\bm k}} \sum_{s=\pm}|r^{cv}_{x,{\bm k},s}|^2 \varepsilon_{{\bm k},s}^{cv} f^{cv}_{{\bm k},s},
\end{align}
where ${\Omega}^{c,{\rm spin}}_{{\bm k}} = {\Omega}^{c}_{{\bm k},\uparrow} - {\Omega}^{c}_{{\bm k},\downarrow}$.
Therefore, the third-order Hall conductivity yields
\begin{align}  
     &\sigma_{yxxx}^{(3),P} (\omega, \omega, -2\omega) \approx C (\omega)
      \sum_{{\bm k}}
    {\Omega}^{c,{\rm spin}}_{{\bm k}} \sum_{s=\pm} |r^{cv}_{x,{{\bm k},s}}|^2 \varepsilon_{{\bm k},s}^{cv} f^{cv}_{{\bm k},s},
\end{align}
where $C(\omega) = e^4/ 2\hbar^4\omega^3$ for charge conductivity and $C(\omega) = e^3/ 2\hbar^4\omega^3$ for spin conductivity. This is the paramagnetic part of the third-order conductivity expression for the contribution at below gap. On the other hand, the diamagnetic part of the spin conductivity in the regime below the band gap yields
\begin{align}
    &\sigma_{yxxx}^{(3),D} (\omega, \omega, -2\omega) \approx \frac{C(\omega)\hbar^2}{e^2}
     \sum_{{\bm k}}
    {\Omega}^{c,{\rm spin}}_{{\bm k}} \sum_{s=\pm} \kappa^{cv}_{xx,{\bm k},s} f^{cv}_{{\bm k},s}. 
\end{align}
The total third-order spin conductivity thus follows 
\begin{align}
    &\sigma_{yxxx}^{(3)} (\omega, \omega, -2\omega) \approx C(\omega)
     \sum_{{\bm k}}
    {\Omega}^{c,{\rm spin}}_{{\bm k}} \sum_{s=\pm}\bigg\{ |r^{cv}_{x,{{\bm k},s}}|^2 \varepsilon_{{\bm k},s}^{cv}  + 
     \tilde{\kappa}^{cv}_{xx,{\bm k},s} \bigg\}f^{cv}_{{\bm k},s}, 
\end{align}
where $\tilde{\kappa}^{cv}_{xx, {\bm k},s} = (\hbar/e)^2
     \kappa^{cv}_{xx,{\bm k},s}$. This derives the equation~\eqref{eqn:SpinCond}.
Therefore, as expected, the nonlinear spin Hall conductivity depends on the Berry curvature.

\end{document}